\documentclass[aps,prl,twocolumn,showpacs]{revtex4}

\usepackage{graphicx}

\def\pra{Physical Review A}

\usepackage{amssymb}
\usepackage{amsmath}

\newcommand{\footnoteremember}[2]{
  \footnote{#2}
  \newcounter{#1}
  \setcounter{#1}{\value{footnote}}
}

\newcommand{\footnoterecall}[1]{
  \footnotemark[\value{#1}]
}

\begin{document}

\title{General conditions for quantum adiabatic evolution}

\author{Daniel Comparat}

\affiliation{                    
Laboratoire Aim\'{e} Cotton, CNRS, Univ Paris-Sud, B\^{a}t. 505, 91405 Orsay, France}

\begin{abstract}
Adiabaticity occurs when, during its evolution, a physical system remains in the instantaneous eigenstate of the hamiltonian. Unfortunately, existing results, such as the quantum adiabatic theorem based on  a slow down evolution ($H(\epsilon t)$, $\epsilon \rightarrow 0$),  are insufficient to describe an  evolution driven by the hamiltonian $H(t)$ itself. Here we derive general  criteria and exact bounds,  for the state and its phase, ensuring an adiabatic evolution for any hamiltonian $H(t)$. As a corollary  we demonstrate that the commonly used condition of a slow   hamiltonian variation rate, compared to the spectral gap,  is indeed sufficient to ensure adiabaticity but only when the hamiltonian is  real and non oscillating (for instance containing exponential or polynomial  but no sinusoidal functions).
\end{abstract}

\pacs{03.65.Ca, 03.65.Ta, 03.65.Vf}

\date{\today}

\maketitle


The ''adiabatic'' process, from the Greek $\alpha-$ (a-), not, $\delta \iota \alpha$ (dia), through, $\beta \alpha \iota \nu \iota \nu$ (bainen), to pass, was introduced by Carnot (in 1824) and W. J. M. Rankine (in 1858) in thermodynamics, then by Boltzmann (in 1866) in classical mechanics\cite{Laidler}.  
 In 1928, Fritz London applied adiabatic process in chemical kinetics.  Concerning the quantum physics,
 in 1911-1916 Paul Ehrenfest used adiabatic invariance in the development of the 'Old Quantum Theory' and in 1928 Born and Fock \cite{1928ZPhy...51..165B} demonstrated the  quantum adiabatic theorem.  By definition,
 quantum adiabaticity occurs when, during its evolution driven by an hamiltonian $H(t)$, a  quantum state $|\Psi(t)\rangle$ prepared in an eigenstate $  | n(0) \rangle $ remains close to the instantaneous eigenstate $| n(t) \rangle $ (with a proper phase choice)  as time $t$ goes on. 
  The basic concept of adiabaticity in quantum theory
has been widely applied in both theories and experiments. Applications
range from energy level
crossings, such  as Landau-Zener transition, Born-Oppenheimer molecular  coupling, collisional processes, 
quantum control or adiabatic quantum computation \cite{Nakamura2002,Teufel2003}.  
Unfortunately,
even for the two-level system,
 no sufficient conditions are known to efficiently describe an adiabatic evolution driven by a general hamiltonian $H(t)$ \cite{2008PhRvL.101f0403D}.
For instance,
 an example as simple as the Schwinger's hamiltonian (solved hereafter) \cite{1937PhRv...51..648S}
$H(t) = \frac{\hbar \omega_0}{2}  \left( \begin{smallmatrix} \cos \theta  & \sin \theta e^{-i  \omega t} \cr 
\sin \theta e^{i  \omega t} & - \cos \theta
\end{smallmatrix} \right)
$, proves that neither
 the ''usual'' adiabatic phase evolution  $ \int_0^t  E_n / \hbar - i  \langle  n | \dot  n \rangle $ nor the 
 commonly used 
approximate adiabatic criterion\footnoteremember{foot_gap}{$ \langle  m  | \dot n \rangle  =
 \frac{\langle m | \dot H | n \rangle }{E_n - E_m }$ comes from the time derivative of $\langle m  | n \rangle =\delta_{m n} $ and of $\langle m | H | n \rangle =  E_n \delta_{m n}$,  where $\delta_{m k}$  is Kronecker's delta.} 
 \cite{Messiah1959}:
\begin{equation}
 \sum_{m \neq n} \frac{ \hbar }{|E_n -E_m |}  \frac{|\langle m |\dot H |  n \rangle| }{  |E_n -E_m | } = 
\sum_{m \neq n} \left| \frac{ \hbar \langle m  | \dot  n \rangle }{E_n -E_m } \right | \ll 1 ,\label{adia_crit}
\end{equation} 
are sufficient (or necessary) to ensure adiabaticity.  
 This statement may look surprising \cite{2004Sarandy,2005PhRvA..72a2114W,2005PhRvL..95k0407T,2006PhRvL..97l8901D,2006PhRvL..97l8902M,2006PhRvL..97l8903M}  
 but is presented in textbooks \cite{Schiff1949,Bransden1989}. 
 It is indeed well known in NMR or in quantum optics (through Rabi oscillation)  that resonant terms can lead to population  transfer, i.e. to a non adiabatic behavior. This is linked to branch points, connecting the different eigenstates of the adiabatic Hamiltonian, and explains for instance that non adiabatic behaviour exist when several successive transitions between pairs of levels occurs \cite{1991PhRvA..44.4280J,1994PhRvA..49.2217N,Stenholm, 2000PhRvA..61f2104W}. Thus, condition  (\ref{adia_crit})is not valid globally.

 It is therefore important to derive  general conditions, for a system and its phase evolution,  which ensure adiabaticity.   This is the goal of this article.  As a corollary we will  answer
 the still pending (even in the two-level case) question: 
why and when the standard condition (\ref{adia_crit}),  
of a slow   hamiltonian variation rate, compared to the frequency associated to the spectral gap $\Delta E_n = \min_{m\neq n} |E_m - E_n |$,  
is  a
sufficient  adiabatic condition. Indeed, we show that condition (\ref{adia_crit})
is  sufficient to ensure adiabaticity but only when the hamiltonian is  real and non oscillating.

  Because almost all
 existing results, as the adiabatic criterion (\ref{adia_crit}) are based on the so called adiabatic limit of a slow down evolution, we shall first start by studying the standard results and by explaining why the standard adiabatic theorem can not help to solve the problem. Hopefully this part will also clarify the recent debate 
concerning the adiabatic phase and adiabatic criterion
\cite{2005quant.ph..9083Y,2006PhRvL..97l8901D,2006PhRvL..97l8902M,2006PhRvL..97l8903M,2006PhRvA..73d2104M,2006PhLA..353...11V,2007PhRvL..98o0402T,2007PhLA..368...18Y,2005PhLA..339..288T,2007PhRvA..76d4102M,2007PhRvA..76b4304W,2008PhRvL.101f0403D,2008PhRvA..77c2109Z,2008PhRvA..77f2114W,2008PhRvA..78e2508R}  following the (over-subtle) ''rediscovery'' by Marzlin and Sanders \cite{2004PhRvL..93p0408M} that condition (\ref{adia_crit}) is not a sufficient one. We shall then derive exact bounds for adiabaticity. We then discussed their validity in a the general two-level case and their simplification in the case of a non oscillating hamiltonian. For clarity some lengthly  calculations are reported in an appendix.

\section{Standard results}

\subsection{Quantum adiabatic theorem}

The Born and Fock's  quantum adiabatic theorem
 has been rigorously demonstrated, several times and by several different methods 
 (see for instance \cite{1928ZPhy...51..165B,Messiah1959,Teufel2003,2007JMP....48j2111J} and references therein), extended  to the infinite dimensional setting by Kato \cite{1936JPSJ....5..435K}, studied as a
geometrical holonomy evolution by Berry \cite{1984RSPSA.392...45B}, extended
to degenerate cases (without gap condition) \cite{Avron} and to open quantum system   \cite{2005PhRvA..71a2331S}.

  In the
non degenerate ($E_m \neq E_n$) case, the adiabatic theorem stipulates  that: 
\begin{equation}
|\Psi_\epsilon (t)\rangle - e^{-i \int_0^t \left(  E_n / \hbar - i  \left\langle  n_\epsilon \left| \dot  n_\epsilon \right\rangle \right.  \right)} |  n_\epsilon(t) \rangle = O(\epsilon) \xrightarrow[\epsilon \rightarrow 0]{} 0,
\label{adia_phase}
\end{equation}
where
 evolution speed is
controlled by
 $\epsilon$ and the subscript stands for the $H_\epsilon (t) =  H(\epsilon t)$  evolution\footnoteremember{foot_Hs}{The path parametrization  is  here 
$s(t)=t/T=\epsilon t \in [ 0,1 ]$ and  $  H_\epsilon (t) =  H(s(t)) $ where $T$ is the evolution time.
The Schr\"odinger equation is then
$i \hbar \frac{ d | \Psi_\epsilon (t) \rangle}{d t} = H_\epsilon(t)   | \Psi_\epsilon (t) \rangle$ or
$i \hbar \epsilon \frac{ d | \Psi(s) \rangle}{d s} = H(s)   | \Psi (s) \rangle$.
More generally a local control of the speed is possible:
 $  H_s (t) = H(s(t)) $  by using a \emph{monotonic} function 
$s(t) \in [ 0,1 ]$  \cite{2002PhRvA..65d2308R}.  An example is the 
interpolating hamiltonian
$H(s) =H_{\rm in} (1-s)  + H_{\rm fin} s $.}. Here the dot designates the time derivative and  $\int_0^t f = \int_0^t f(t') d t'$.

To illustrate the limited practical utility of the theorem,
let's suppose that an external laser  field, with constant angular frequency $R_1 (t) = \omega$, is applied to a two-level system that
we want to adiabatically drive
 by experimentally modifying
 two parameters: 
 the coupling Rabi frequency (proportional to the square-root of the laser intensity)  $R_2 (t) = \Omega(t) $, and  the detuning of the laser from resonance $R_3(t) = \delta (t)$.
 The hamiltonian is, in the rotating wave approximation:
$
  \frac{\hbar }{2}\left( \begin{smallmatrix} \delta(t) + \omega  & \Omega(t) e^{ - i   \omega t} \cr 
\Omega(t) e^{ i  \omega t} & - \delta(t) - \omega  \end{smallmatrix} \right) =  H (R_1(t),R_2(t),R_3(t),R_4(t))
$.
Due to the $ R_4(t) = \omega t$ term, slowing down the time would lead to $ \omega ( \epsilon t) = (\omega \epsilon) t$. When $\epsilon \rightarrow 0$, this would require  reducing  $\omega$  to zero  which is experimentally impossible.
  Moreover, even in the static field  ($\omega=0$) regime,  the theorem applies but only
   if $\delta$ and $\Omega$  can be  slowed down simultaneously. 
  The theorem
    says nothing  about the adiabaticity if $\delta(t)$ and $\Omega(t)$ are varied independently with time. 
   
     Although undoubtedly of great theoretical interest, as in  the quantum adiabatic computation 
 using interpolating hamiltonian 
 \cite{2001Sci...292..472F}, the theorem describes an
  evolution driven by $H(\epsilon t)$ with  $\epsilon \rightarrow 0$ and is obviously 
 of no utility concerning the evolution driven by $H(t)$ itself, as in this case $\epsilon =1$ and cannot be reduced to zero.
 The theorem  is then better formulated within the parameter domain than within the time domain \cite{2005PhRvA..72a2114W,2007PhLA..370...17L}: 
  an evolution driven by  $  H({\mathbf R}(t)) $  is adiabatic if the parameter path, 
   between an initial ${\mathbf R}_{\mathrm{in}}$ parameter value and a final one ${\mathbf R}_{\mathrm{fin}}$,
  is followed infinitely slowly.

 \subsection{Approximate adiabatic condition}

Contrary to
the quantum adiabatic theorem, the approximate adiabatic condition (\ref{adia_crit}) can be applied to $H(t)$ itself.
 The origin of condition (\ref{adia_crit})
arises \cite{Messiah1959} from the fact that the error term in Eq. (\ref{adia_phase}) can be written \cite{2007JMP....48j2111J} as 
$
O(\epsilon)  =  \sum_{m \neq n} \frac{ \hbar \left\langle  m_\epsilon \left| \dot n_\epsilon \right\rangle \right.  }{E_n -  E_m } + O(\epsilon^2),
$ where
the linear $\epsilon$ dependence is here only implicit and, deliberately but confusingly, hidden  in $\left| \dot n_\epsilon \right\rangle$. This
 has been the source of confusion\cite{2004Sarandy,2005PhRvA..72a2114W,2005PhRvL..95k0407T,2006PhRvL..97l8901D,2006PhRvL..97l8902M,2006PhRvL..97l8903M}  when used with  $\epsilon=1$ where $ | m_\epsilon \rangle = | m \rangle$.
 The confusion occurs
 because, even if derived without any proof  by using $\epsilon=1 \nrightarrow 0$,
the criterion  (\ref{adia_crit}) 
 ensures an adiabatic evolution in almost all the known examples: Landau-Zener(-St\"uckelberg), Rosen-Zener-Demkov, Nikitin, Zhu-Nakamura models or in the
Rapid Adiabatic Passage  or  STImulated Raman Adiabatic Passage (STIRAP) processes, ...
 \cite{Nakamura2002, Teufel2003, Nikitin2006}. 
 Important enough, as we shall see, all these examples use non-oscillating (exponential or polynomial) functions. 
  Therefore,
the simple idea of adiabaticity, given by the condition (\ref{adia_crit}),
of a
small but finite  variation rate of $H(t)$ (compared to the spectral gap), 
 is  broadly  used. 
 Similarly, as extracted from equation (\ref{adia_phase}) without any proof  by using $\epsilon=1 \nrightarrow 0$, an adiabatic phase evolution of $ \int_0^t  E_n / \hbar - i  \langle  n | \dot  n \rangle $ is widely used \cite{2005PhLA..339..288T}.
 However  as mentioned in the introduction 
the Schwinger's example 
 demonstrates that this ''usual'' adiabatic condition, as well as this ''usual'' adiabatic phase,
is neither sufficiently nor necessarily  to obtain an adiabatic evolution.

To avoid any confusion the term $O(\epsilon^2)$ has to be evaluated. This can be done for instance by giving an exact bound \cite{2007JMP....48j2111J} on the adiabatic fidelity $|\langle \Psi | n \rangle |$ such as  \footnoteremember{foot_norm}{In this article we use $\|x\| = \sqrt{x^\dag x} =  \sqrt{\sum_{i=i}^N |x_i|^2} $ and $\|x\|_1 = \sum_{i=1}^{N} |x_i| $. They verify $\|x\|_2\leq \|x\|_1 \leq \sqrt{N} \|x\|_2$.
 For the matrix norm $\|M\| = \max_{x\neq 0} \frac{\| M x\|  }{\|x\|}$.}
\begin{eqnarray}
 \frac{1- |\langle \Psi (t) | n (t) \rangle |}{\hbar} & \leq  &
 \left\| \frac{\dot H (0)}{{\Delta E_n (0) }^2} \right\|  +  \left\| \frac{\dot H (t)}{{\Delta E_n (t) }^2} \right\| + \nonumber \\
 & & 
   \int_0^t \! \! \! \left(  \frac{7 {\| \dot{ H} \|}^2}{{\Delta E_n}^3} +  \frac{\| \ddot{ H} \|}{{\Delta E_n}^2} \right).
   \label{bound}
   \end{eqnarray}
  Similar bounds \cite{Teufel2003,2006PhRvA..73f2307S,2007PhRvL..98o0402T} exists. They can be used to restore the usual theorem\footnoterecall{foot_Hs} because 
 $ \frac{d H_\epsilon}{dt } = \epsilon \frac{d H_\epsilon}{d(\epsilon t)} = \epsilon \frac{d  H}{d s} $ vanishes when $\epsilon  \rightarrow 0$. However, they
 have severe limitations because, due to the integral term, 
they  require a maximal evolution time $T$  to provide an adiabatic evolution when none is needed. This can be easily seen from the 
Schwinger's 
$H(t) = \frac{\hbar 10 \omega}{2}  \left( \begin{matrix} \cos \theta  & \sin \theta e^{-i  \omega t} \cr 
\sin \theta e^{i  \omega t} & - \cos \theta
\end{matrix} \right)$ example with $\omega=1\,$s$^{-1}$ and $\theta =0.01$.

  \section{General bounds}
  
In order to derive a more useful bound than (\ref{bound}), let's study
the evolution of $|\Psi (t) \rangle$ driven by a general $N$-level hamiltonian $H(t)$. For the corresponding eigenvalues $E_m(t)$ of $H(t)$,
the eigenvectors $e^{i \theta_m(t)} |m(t)\rangle$, $m=1,\cdots,N$ form  a so called adiabatic basis,
where $\theta_m(t)$ are arbitrary phases to be chosen conveniently later.
  To study the adiabatic evolution we assume that $| \Psi(t=0) \rangle = | n (0)\rangle$. 
  The  Schr\"odinger equation for $|\Psi (t) \rangle = \sum_{m=1}^N U_{mn}(t)  e^{i \theta_m(t)} |m (t)\rangle $, i.e. with 
 $U(0)=I$, leads to  
  the time-evolution equation:
$ i \hbar \dot  U   = H'  U 
$, where
\begin{equation}
 H'_{m k} = (E_m +\hbar \dot \theta_m )\delta_{m k} - i \hbar \langle m | \dot k \rangle e^{i(\theta_k-\theta_m)}
 \label{Had_def}.
 \end{equation}
 As usual, we identify the operators and their matrices in the standard (also called natural or canonical) basis
  $|m^{\rm st} \rangle$, $m=1,\cdots,N$. Thus,  $H' = P^{-1} H P - i \hbar  P^{-1} \dot  P $ where
the columns of $P$ are the eigenvectors $e^{i \theta_m(t)}  |m\rangle$: $P_{m k} =  \langle m^{\rm st} | P |k^{\rm st} \rangle = e^{i \theta_k(t)} \langle m^{\rm st} |k \rangle $ of $H$.

The evolution is adiabatic if and only if the fidelity
$ |U_{n n} (t)  |$ is close to unity  (or $\|  |\Psi \rangle \langle \Psi |- | n \rangle \langle n | \| \ll 1 $).   In order to also study the phase evolution of the state $|\psi\rangle$ 
we compare the matrix $U$ to  another time evolution matrix $  U'$ which can be  more easily evaluated. 

Let's define $U'$  by $U'(0)=1$   and
 \begin{equation}
 i \hbar \dot U' = P'^{-1} H' P' U',  \label{Schro_Uprime}
\end{equation}
where $P'$ is an auxiliary matrix to be chosen conveniently.
Then, the important equality comparing two operators
\begin{eqnarray}
 U(t) - U'(t) & = & (P'(t)-1)U'(t) - U(t)(P'(0)-1)  -     \nonumber \\
  & & U(t) \int_0^t U^{-1}(t') \dot{P'}(t') U'(t') d t' \label{bound_U_Uprime}
\end{eqnarray}
can  be established by multiplying it by $U^{-1}$ and then taking the time derivative. 

Several choices are possible but, for simplicity we choose $P'$  to have $P'^{-1} H' P'$ as an eigenvalue value decomposition of $H'$.
In this case $U'$ is diagonal:
$U'_{ n n} = e^{-i \int_0^t E_n'(t')/\hbar \ d t'}$
where $E_n'$ is the eigenvalue  of the $n^{\rm th}$  eigenvector $|n'\rangle=P'|n^{\mathrm{st}}\rangle$ of $H'$.
$P'$ is unitary, so $\| P' \| = \| U \| = \| U' \| = 1 $. We then apply
Eq. (\ref{bound_U_Uprime}) on  $|n^{\mathrm st} \rangle$ and take the norm on both sides 
to have
\begin{eqnarray}
\lefteqn{ \left\| |\Psi (t) \rangle - e^{-i \int_0^t \frac{E_n'}{\hbar} } |n (t) \rangle \right\|   \leq 
\left\| | n' (0) \rangle - |n^{\rm st} \rangle   \right\| + } \nonumber \\
&  &   \left\| | n'  (t) \rangle - |n^{\rm st} \rangle   \right\|+ \int_0^t \! \! \left\| | \dot{n'} \rangle  \right\|
\label{key_result}
\end{eqnarray}
Eq. (\ref{key_result}) gives
a  bound
as well as 
the correct phase evolution for adiabatic evolution. 
Sufficient adiabatic conditions are: 
\begin{eqnarray}
 \lefteqn{ \left\| (P'(t) - 1) |n^{\rm st} \rangle  \right\| = \left\| | n' (t) \rangle - |n^{\rm st}  \rangle \right\|   \ll  1 } \label{first_th1} \\
 & & \int_0^t \left\|  \dot{P'} (t')  |n^{\rm st} \rangle   \right\| dt' = \int_0^t \left\| | \dot{n'} (t') \rangle  \right\| dt' \ll  1 \label{first_th2}
\end{eqnarray}
To tight these bounds we choose the phase of $|n'\rangle$ to be such that $\langle n^{\rm st} | n' \rangle \geq 0$.  The adiabatic fidelity is bound by the inequality $2 (1-| \langle \Psi(t)|n(t) \rangle|) \leq \| |\Psi (t) \rangle - e^{-i \int_0^t \frac{E_n'}{\hbar} } |n (t) \rangle \|^2  $, which
 should now be tighten as much as possible by choosing the $\theta_m(t)$ phases.

 Links and differences, of Eq. (\ref{key_result}) with the usual theorem  given by Eq. (\ref{adia_phase}) and of
 Eq. (\ref{first_th1}) with the usual condition given by Eq. (\ref{adia_crit}), can be inferred by applying standard  perturbation theory  to Eq. (\ref{Had_def}):
  \begin{eqnarray}
E'_n  & \approx  & E_n - i \hbar \langle n | \dot n \rangle + \hbar \dot \theta_n +  \sum_{m \neq n} \frac{|H'_{m n}|^2}{H'_{nn} - H'_{mm}}  \\
 | n' \rangle & \approx  & |n^{\rm st} \rangle + \sum_{m \neq n} \frac{H'_{m n}}{H'_{nn} - H'_{mm}} |m^{\rm st} \rangle. \label{stand_pert_the}
  \end{eqnarray}
 Using the equality
 $ \theta_m = \theta_n + \arg (-i \langle m | \dot n \rangle)  $ for all $m\neq n$, creates (to this second order approximation) reals $P'_{m n}$, and condition  (\ref{first_th1}) becomes:
\begin{equation}
 \sum_{m \neq n} \! \! \left| \frac{\langle m | \dot n \rangle }{  (E_n-E_m)/\hbar - i  \langle n | \dot n \rangle  + i \langle m | \dot m \rangle
 - \frac{d}{d t} \arg \langle m | \dot n \rangle} \right| \! \! \ll \! \! 1  \label{cond_pert} 
\end{equation}
This condition, first derived in  \cite{2005quant.ph..9083Y}, generalizes  condition (\ref{adia_crit}) when $H$ is not real\footnote{For real hamiltonian $H$, $|m\rangle$ is real, so the (Pancharatnam's phase) $\arg \langle m | \dot n \rangle$ is zero, and  $\frac{d}{dt} \langle m | m \rangle  = 2 \langle m | \dot m \rangle = 0$.}.
 However, as (\ref{adia_crit}), it is  an insufficient adiabatic criterion for two reasons: it arises from a perturbative approach, and it neglects the 
condition (\ref{first_th2}), which is important for oscillating $H$. Indeed,   
it is only when the hamiltonian matrix elements are non oscillating functions --
in the approximate sens of none of their sum, product, division or combination has a large number of monotonic changes --
that the condition (\ref{first_th2}) can be neglected. More precisely, in the general case
when 
all the $ P'_{m n} = \langle m^{\mathrm st} | n' \rangle $ are real (or with a time independent phase argument) and monotonic,  an important simplification occurs because 
$\int_{0}^{t} | \dot P'_{m n}(t') | d t' =  |P'_{m n}(t)- P'_{m n}(0)| \leq |P'_{m n}(t)-1| +| P'_{m n}(0)-1|  $. In this case, we see,  by using the 1-norm\footnoterecall{foot_norm}, that  the derivative condition (\ref{first_th2})  essentially reduces to the sole (\ref{first_th1}) condition. Similarly $P'_{m n}$ piecewise  functions with finite number ($M-1 $) of monoticity changes\footnote{Interestingly enough, very similar considerations have been used by Born and Fock in their seminal paper \cite{1928ZPhy...51..165B}.} would lead to a
 $ \sum_m  |P'_{m n}-1| \ll  1/M$ type of condition.

      \subsection{Multi levels system}
      
We 
 use here an exact perturbation theory  \cite{Paldus2006} to calculate $P' |n^{\mathrm{st}} \rangle$.
We write
$H' = H_0 + V $ where $V$ is a perturbation. For simplicity, i.e.
 in order to isolate the $n^{\mathrm{th}}$ subspace, we renumber the states to have $n=1$ and, using the $1+(N-1)$ block matrix notation, we choose (see Eq. (\ref{Had_def})) 
$$
H' = H_0 + V ; \ \ 
  H_0 = \left( \begin{smallmatrix} H'_{n n} & 0  \cr 
0   &   H'_{n n} - \hbar \delta'
\end{smallmatrix} \right) , \  \ V = \frac{\hbar}{2} \left(  \begin{smallmatrix} 0 &  \Omega'^\dag  \cr 
\Omega'    &   0
\end{smallmatrix} \right).
$$
We then apply techniques, detailed in the appendix,  to endup with the following simple conditions:
\begin{eqnarray}
\| \delta'^{-1} \| \| \Omega' \| & \ll & 1 \label{fourth_th} \\
  \int_0^t \left( \| \Omega' \|  \left\| \frac{d}{dt} (\delta'^{-1})  \right\|  +  \| \delta'^{-1} \|  \left\| \frac{d}{dt}  \Omega'  \right\| \right)  & \ll & 1  \label{fourth_th2} 
 \end{eqnarray}
 which are together  sufficient adiabatic conditions because they imply  Eqs. (\ref{first_th1}) and (\ref{first_th2}).

 Eq. (\ref{fourth_th2}) is here to prevent the use of oscillating hamiltonian. Indeed,
 as discussed previously, if $H$ is real and ''non-oscillating'', meaning that $\Omega'_{m n}$ and $(\delta'^{-1})_{m k}$ are (piecewise) real monotonic functions, the
 condition (\ref{fourth_th2})  essentially reduces to  condition (\ref{fourth_th}).

Eq. (\ref{fourth_th}) itself can be seen as a generalization of the Eq. (\ref{cond_pert})  which itself generalizes the standard condition (\ref{adia_crit}).
Indeed, 
 if we add to the condition (\ref{fourth_th}) the, fortunately common condition of  negligible coupling within the space orthogonal to $|n\rangle$, i.e. negligible\footnote{Simple condition (similar to Weyl's theorem), for the smallness of ${\delta '}^{-1}$ off diagonal terms exists. For instance by choosing $H_0$ as the diagonal part of $H'$ in the Brillouin-Wigner equation we can replace conditions (\ref{fourth_th}) essentially by $	\frac{\| \dot H (t)   \|_1}{ \Delta E(t)} \ll \frac{{\Delta E}_{n} (t)}{\hbar} $. However, this may be of small practical interest because it contains  not only  ${\Delta E}_{n} $, the  gap relative to $|n\rangle$, as in Eq. (\ref{adia_crit}), but also  the global energy spectral gap $\Delta E = \min_{m \neq k} |E_k -E_m |$. } $\delta'$ off diagonal terms to have $({\delta'}^{-1})_{mm} \approx ({\delta'}_{mm})^{-1}$, we can recover Eq. (\ref{cond_pert})
by choosing
$\theta_m = \theta_n + \arg  (-i \langle m | \dot n \rangle)  $ (i.e. $\Omega'$ real).

 Finally, the appendix indicates that,
  for a 
  strongly  non-oscillating (very few monotonicity changes) real hamiltonian $H$  
  the sole usual condition (\ref{adia_crit}), which is 
  then condition (\ref{fourth_th}),
   is sufficient to ensure an adiabatic behavior.

     \subsection{Two level system}
     
 Let's now illustrate the results in the two-level ($N=2$) framework. We write, by removing the average diagonal energy,  the general hamiltonian in the (spin-magnetic interaction $H=- \gamma \vec B.  \frac{\hbar}{2} \vec \sigma$) form:
 $H(t)  =   \frac{\hbar \omega_0(t)}{2}  \left( \begin{smallmatrix} \cos \theta(t)  & \sin \theta(t) e^{-i \varphi(t)} \cr 
\sin \theta(t) e^{i \varphi(t)} & - \cos \theta(t)
\end{smallmatrix} \right)
$. 
Using $-\theta_1  =\theta_2= \theta_1 +\arg( - i \langle 2|\dot 1 \rangle)$,
the appendix   shows that
 Eq. (\ref{first_th1}) and Eq. (\ref{first_th2}) are equivalent to:
\begin{eqnarray}
\frac{|\dot \varphi \sin \theta - i \dot \theta|}{\left|\dot \varphi \cos \theta - \omega_0 - \frac{d}{d t} \arg ( \dot \varphi \sin \theta - i \dot \theta)  \right|}  = \frac{|\Omega'|}{|\delta'|}  \! \! & \ll & \! \! 1 \label{two_level_cond1}\\
\int_0^t dt' \left| \frac{d}{dt'}\frac{|\Omega'(t')|}{\delta'(t')} \right| \! \!  & \ll & \! \!  1 \label{two_level_cond2}
\end{eqnarray}

In order to check their validity, or similarly the one of conditions  (\ref{fourth_th}) and  (\ref{fourth_th2}), we first use the simple example due to Schwinger \cite{1937PhRv...51..648S},
where all the parameters $\omega_0,\theta,\dot \varphi=\omega$ are real and time independent. In this case the
 condition (\ref{two_level_cond2}) vanishes and
 $U(t) = e^{ -i\int_0^t H'/\hbar}=
\left( \begin{smallmatrix}
(\cos\frac{\Omega_R t}{2}-i\frac{\delta'}{\Omega_R}\sin\frac{\Omega_R t}{2} ) & -i \frac{\Omega'}{\Omega_R}\sin\frac{\Omega_R t}{2}  \\
-i  \frac{\Omega'}{\Omega_R}\sin\frac{\Omega_R t}{2}
&( \cos\frac{\Omega_R t}{2}+i\frac{\delta'}{\Omega_R}\sin\frac{\Omega_R t}{2} )  \end{smallmatrix} \right)
$
where $\Omega_R = \sqrt{|\Omega'|^2+\delta'^2} $ is
the generalized Rabi frequency. 
  The adiabatic evolution (negligible off-diagonal terms in $U$) is ensured by
 the condition
$
  \frac{|\Omega'|}{\Omega_R} = \frac{|\omega \sin \theta|}{\sqrt{(\omega_0-\omega\cos\theta)^2 + \omega^2 \sin^2\theta}} \ll 1
$ which is indeed equivalent to our condition (\ref{two_level_cond1}): $
  \frac{|\Omega'|}{|\delta'|} = \frac{|\omega \sin \theta|}{|\omega_0-\omega\cos\theta|} \ll 1
$. Furthermore, our equation (\ref{key_result}), including its phase $ \int_0^t   E'_1 / \hbar  = \Omega_R t /2$, correctly describes an adiabatic evolution.

On the contrary, using
this analytical example
(by looking at the resonant $\omega \approx \omega_0$ or small $\theta$ cases for instance) it is straightforward to demonstrates 
 that  Eq. (\ref{adia_crit}): $\frac{|\Omega'|}{|\omega_0|} = \frac{|\omega \sin \theta|}{|\omega_0|}  \ll 1$, as well as the ''usual'' adiabatic phase evolution  $\omega_0 t/2 = \int_0^t  E_1 / \hbar - i  \langle  1 | \dot  1 \rangle $ (see Eq. (\ref{adia_phase})),
are not correlated with an adiabatic evolution.

We now add to our study the  condition (\ref{two_level_cond2}) by the use of the real cycling hamiltonian $H =  \frac{\hbar}{2}
 \left( \begin{smallmatrix}
\delta  & \Omega \cr 
 \Omega & - \delta\end{smallmatrix} \right)$
where $\delta(t) =  \alpha \cos (\varpi t)$ and $\alpha,\varpi,\Omega$ are positive constants verifying, for simplicity,  weak-coupling ($ \Omega \ll \alpha$) and large amplitude ($\alpha \gg \varpi$).
For $t\in \left[ 0 , T_1 =   \pi / \varpi \right ]$,  
$  \frac{|\Omega'|}{\delta'}  = \frac{|\dot \theta|}{\omega_0} =  \frac{ \dot \Omega \delta -\Omega \dot \delta }{(\delta^2 + \Omega^2 )^{3/2}} $ is real and with a single monotonicity change, so our second condition (\ref{two_level_cond2}) reduces to condition (\ref{two_level_cond1}). 
The non-adiabatic transition probability $p_1$
(so called single-passage or one-way  transition),
 is given by  the
Landau-Zener's formula: $p_1 \approx e^{- \frac{\pi}{2} \frac{\Omega^2}{ \alpha \varpi} }  $  \cite{1994PhRvA..50..843K} and
 the adiabatic limit $p_1\rightarrow 0$  is covered by the condition (\ref{two_level_cond1}): $\max_{t\in [ 0 , T_1   ] }
\left| \frac{\dot \theta}{\omega_0} \right| =
\frac{\alpha \varpi }{\Omega^2}    \ll  1 $.  
After $M$ (even) multiple passage, for $t= M T_1 $,  the  non-adiabatic transition probability  becomes $p_{M}  \approx p_1 \frac{\sin^2 M\Theta}{\cos^2 \Theta} $ and depends of a relative  (St\"uckelberg) phase  $ \Theta \simeq \frac{\alpha}{\varpi}$ 
 of the wavefunction\cite{1994PhRvA..50..843K}.
 For $ \Theta \sim \pi/2 [\pi]$, $p_M$ can be  $M^2$ times higher than $p_1$ leading to a full non adiabaticity $p_{M} \sim 1$ even if $p_1 \ll 1$. This
 illustrates why, in such an oscillating case, 
 condition (\ref{two_level_cond1}) ($\frac{\alpha \varpi }{\Omega^2}    \ll  1 $) is not sufficient and the extra
 condition (\ref{two_level_cond2}) ($ \frac{\alpha \varpi }{\Omega^2}    \ll  1/M $)  is  needed to ensure an adiabatic evolution.
  This example shows that, with an oscillating hamiltonian, even if a single passage is quasi-adiabatic constructive  interferences might accumulate  the small non adiabatic amplitude to result, after multiple passages, in a full non-adiabatic transition\footnote{Interestingly enough, 
the reverse case, namely the diabatic limit ($p_1 \approx 1$) can lead (for instance when  $ \alpha/\varpi $ annul the Bessel $J_0$ function) to the reverse phenomenum of adiabaticity created after  multiple passages ($p_M \approx 0$) 
known as suppression of the tunneling,  coherent destruction of tunneling, 
 dynamical localization or  population trapping depending on the context   
\cite{Grifoni,1994PhRvA..50..843K}.}. 
  This is very similar to the case of single crossing but with several levels \cite{1991PhRvA..44.4280J,2004AcPPB..35..551G}, or to multilevel system \cite{2000PhRvA..61f2104W}, leading, using stationary phase (saddle-point) theorem 
 or steepest descent WKB type of methods, to sums or products of
dephased Landau-Dykhne-Davis-Pechukas's  formulas 
 corresponding to several successive
transitions between pairs of levels \cite{1994PhRvA..49.2217N,Stenholm}.
Finally,  this shows that the standard  condition 
(\ref{adia_crit})
breaks down, not only when resonant terms are present, as sometimes believed  \cite{2006PhRvL..97l8901D,2006PhRvA..73d2104M,2006PhLA..353...11V,2009PhRvL.102v0401A},  
but more generally 
 when oscillating terms are presents.

     \section{Conclusion}

  By simply diagonalizing the hamiltonian $H'$ (hamiltonian in the adiabatic basis), we have derived
simple conditions, Eqs.(\ref{fourth_th}) and (\ref{fourth_th2}) and exact bounds (Eq. (\ref{key_result})) for the state and its phase, ensuring an adiabatic evolution. 
The usual (or standard) condition (\ref{adia_crit})  is found to be a
sufficient adiabatic condition but only for a real and ''non-oscillating'' hamiltonian evolution.   
  This  explains why all the previously cited examples (Landau-Zener, STIRAP, ...) deal with the (real) interaction representation or the dressed state basis, where
 $\omega = 0 $, and  use non oscillating functions such as exponential or polynomial ones.

  Condition  (\ref{fourth_th2}) prevents oscillation\footnote{More practical definition of a ''non-oscillating'' hamiltonian than based on the small number of monotonicity change of the adiabatic hamiltonian $H'$ would be useful. We conjecture (and hope that someone could demonstrate it)  that real hamiltonian containing sum, product, multiplication or division of composition of (real) exponential or polynomial functions are of this type. A possible clue for this proof may be based on iterative 
 used of Eq. (\ref{bound_U_Uprime}) with the 
 iterative  $U'',P'', U^{(3)}, P^{(3)} \cdots$ matrices converging toward the wanted diagonalization, as done in the Jacobi algorithm which iterates the $N=2$ case. During each Jacobi step, the $P^{(m)}$ elements are still of the sum, product, multiplication or division of composition of (real) exponential or polynomial types. The exponent of the polynomial functions grows,  as well as the number of monoticity changes,  but slowly enough  to be always bounded (they never reach infinity due to the isolated zero theorem).} 
   but unfortunately with no distinction between case with constructive crossings or case with destructive (St\"uckelberg) interferences.
  However, the generic most common case concerns a ''complex enough'' system  with small total probability when the single crossing probability is small \cite{Akulin2006}, i.e. where the sole Eq. (\ref{fourth_th}), or Eq. (\ref{adia_crit}) for real hamiltonian, is sufficient  to ensure an adiabatic evolution.

This result simply highlight the fact that
 the standard mathematical technique (so called asymptotic
analysis) to study the
adiabaticity consists in extracting, form the global solution of the Schroedinger equation, a set
of local solutions which individually covers a region (let say between time 0 and $T$), with a controlled
behavior of the coefficients in the equation. This  means that the criterion (\ref{adia_crit}) is local and that
in order to study the adiabatic behavior of a given hamiltonian, one should cut its evolution in part where we could apply safely the criterion (\ref{adia_crit}), namely in part with single branching point or with single crossing between pairs of levels.
Globally we shall add each local non-adiabatic amplitude to get the global non-adiabatic amplitude
\cite{1991PhRvA..44.4280J,2004AcPPB..35..551G,2000PhRvA..61f2104W,1994PhRvA..49.2217N,Stenholm}. 
We would stress that all this should be very well known, but seems to be forgot by many physicist if we refer to recent published articles.
Our article, demonstrate in a simple way that using non-oscillating function the number of local solution is obviously finite and so the added probability remains small if the criterion (\ref{adia_crit}) is globally fulfilled.

 Finally, the adiabatic evolution is  strongly related  to the (semi-)classical limit $\hbar \rightarrow 0$ of quantum mechanics\footnoterecall{foot_Hs} \cite{1984JPhA...17.1225B}, to the WKB approximation \cite{2004AcPPB..35..551G},
to the Minimal work principle \cite{2005PhRvE..71d6107A}, to the quasistatic thermodynamical process \cite{1367-2630-8-5-083},
and to perturbation theory. Therefore, we  hope that
this work and the given examples can enable the development of significant techniques, or
provide novel insights into these important systems.

\acknowledgments
 Thanks to Sabine Jansen to have pointed out to me Born and Fock's consideration concerning monotonicity.

\section{Appendix}

\subsection{Multi levels model}

We demonstrate here that conditions  (\ref{fourth_th}) and (\ref{fourth_th2}) imply
 the conditions (\ref{first_th1}) and (\ref{first_th2}). The techniques are similar to one 
used in Davis-Kahan sin $\theta$ theorem or Weyl-Bauer-Fike's types of perturbative bounds \cite{Ipsen,Chen2006,li:643}.
The starting point is the  exact  Brillouin-Wigner  perturbation theory   that we demonstrate here for completeness \cite{Paldus2006}. 

We define the projector $Q_n = 1 - | n^{\rm st} \rangle \langle n^{\rm st} |$, which in matrix notation is $   Q_n = \left( \begin{smallmatrix} 0 & 0  \cr 
0   &   1
\end{smallmatrix} \right) \ $, and
 the eigenvector  $| \mathbf{n}' \rangle \propto | n' \rangle  $ of $H' = H_0 +V$ with a simple normalization $ \langle n^{\rm st } | \mathbf{n}' \rangle =1$.
$H_0 =  H' - V $ commutes with $Q_n$ so\footnote{We wrote $E'_n 1$ simply as $E'_n$}
  $( E'_n - H_0) Q_n | \mathbf{n}' \rangle = Q_n ( E'_n - H' + V)  | \mathbf{n}' \rangle = Q_n V | \mathbf{n}' \rangle $. When multiply by
  $ (E'_n - H_0)^{-1}$
  this directly lead, using $Q_n | \mathbf{n}' \rangle = | \mathbf{n}' \rangle - |  n^{\rm st}  \rangle  $, to
  the Brillouin-Wigner equation:
$| \mathbf{n}' \rangle   =  (1 - (E'_n - H_0)^{-1} Q_n  V)^{-1} | n ^{\rm st} \rangle$. Using the  matrix notation, and the blockwise inversion,
 this  becomes:
\begin{equation}
 | \mathbf{n}' \rangle   =   \left( \begin{smallmatrix} 1 \cr 
(1+  \delta'^{-1} \Delta'  )^{-1}    \frac{\delta'^{-1} \Omega'}{2} 
\end{smallmatrix} \right)  \label{BW1}
\end{equation}
 where
 $ \hbar \Delta' = E'_n  - H'_{n n} = \langle n^{\rm st} | H_0 + V | \mathbf{n}' \rangle  - H'_{n n} = \langle n^{\rm st} | V | \mathbf{n}' \rangle $  satisfies 
\begin{equation}
\delta'^{-1} \Delta'  =    \frac{\delta'^{-1}  \Omega'^\dag }{2}  (1+  \delta'^{-1} \Delta'  )^{-1}    \frac{\delta'^{-1} \Omega'}{2} 
\label{BW2}
\end{equation}
The idea is now to use the smallness of
 $\delta'^{-1} \Omega'$ (see Eq.  (\ref{fourth_th})) to evaluate $ | n' \rangle =  | \mathbf{n}' \rangle /\sqrt{\langle  \mathbf{n}' | \mathbf{n}' \rangle  }$ and its time derivative, i.e. to study Eqs. (\ref{first_th1}) and (\ref{first_th2}).

We first take the norm of Eq. (\ref{BW2}) and use $\|  (1+  \delta'^{-1} \Delta'  )^{-1} \| \leq  \sum_{k=0}^\infty   \| \delta'^{-1} \Delta'  \|^k  = (1 -  \| \delta'^{-1} \Delta'  \| )^{-1}$ to see that: 
if  $\| \delta'^{-1} \| \| \Omega' \|  \ll  1$ then $\| \delta'^{-1} \Delta' \|  \ll  1$ (see also
Eq. (\ref{WBF})). Therefore Eq. (\ref{BW1}) shows that
Eq. (\ref{first_th1})
is implied by (it is in fact equivalent to)  Eq.  (\ref{fourth_th}): $\| \delta'^{-1} \| \| \Omega' \|  \ll  1$. 

When  $\| \delta'^{-1} \| \| \Omega' \|  \ll  1$, i.e. $| \mathbf{n}' \rangle \approx |  n^{\rm st} \rangle  $, the time derivative of
$ | n' \rangle =  | \mathbf{n}' \rangle /\sqrt{\langle  \mathbf{n}' | \mathbf{n}' \rangle  }$ shows that
 $ \left\| | \dot{n'}  \rangle  \right\| \approx   \left\| | \dot{\mathbf{n}'}  \rangle  \right\|$.  
 Time derivative of the  equations
 (\ref{BW1}) and (\ref{BW2}) can then be used to study Eq.  (\ref{first_th2}). Indeed,
using $\dot{ \delta'^{-1} }= - \delta'^{-1} \dot \delta' \delta'^{-1}$ and (again)
useful estimations on the smallness (see Eq. (\ref{BW2})) of the norm of $ \delta'^{-1} \Delta' $ and its time derivative,
 finally leads to the fact that condition (\ref{fourth_th2}) (together with (\ref{fourth_th})) implies the condition (\ref{first_th2}).

\subsection{Two levels model}
We derive here, in a simpler way, the conditions
  (\ref{fourth_th}) and (\ref{fourth_th2}).
The  eigenvectors $  e^{i \theta_1} |1 \rangle, e^{i \theta_2} |2 \rangle$ of the hamiltonian $H$,
  corresponding respectively to the eigenvalues $ \hbar \omega_0/2$ and $-\hbar \omega_0/2$, are given by the columns of $P= \left( \begin{smallmatrix}
e^{ -i \frac{ \varphi}{2}} \cos \frac{\theta}{2} e^{i \theta_1} & - e^{- i \frac{ \varphi}{2}} \sin \frac{\theta}{2} e^{i \theta_2}   \cr 
  e^{ i \frac{ \varphi}{2}} \sin \frac{\theta}{2} e^{i \theta_1}  & e^{ i \frac{ \varphi}{2}} \cos \frac{\theta}{2} e^{i \theta_2} 
\end{smallmatrix} \right)$
and
$
H' =   \frac{\hbar}{2} \begin{pmatrix} - \dot \varphi \cos \theta + \omega_0  + 2 \dot \theta_1 & ( \dot \varphi \sin \theta + i \dot \theta)  e^{i (\theta_2-\theta_1) } \cr 
( \dot \varphi \sin \theta - i \dot \theta)  e^{i (\theta_1-\theta_2)} &   - \omega_0  + \dot \varphi \cos \theta + 2 \dot \theta_2
\end{pmatrix}.
$

With $\theta_2=-\theta_1$, $H' =   \frac{\hbar}{2}  \left( \begin{smallmatrix} \delta'  & \Omega'^\dag \cr 
\Omega'  & - \delta'
\end{smallmatrix} \right)$.
The $N=2$ case is a very special one because it is always possible to choose $H'$, and then $P'$ real with $\theta_2= \theta_1 +\arg( - i \langle 2|\dot 1 \rangle) $. 
Using the obvious notations 
$ H' =   \frac{\hbar \omega'_0}{2}  \left( \begin{smallmatrix} \cos \theta'  & \sin \theta'  \cr 
\sin \theta'  & - \cos \theta'
\end{smallmatrix} \right)
$,
i.e.
$P'= \left( \begin{smallmatrix}
 \cos \frac{\theta'}{2}  &  - \sin \frac{\theta'}{2}   \cr 
   \sin \frac{\theta'}{2}   & \cos \frac{\theta'}{2} 
\end{smallmatrix} \right)  
$.
Our conditions (\ref{first_th1}-\ref{first_th2}) then read
$\theta'\ll 1$ and $\int_0^t|\dot \theta'| \ll 1$ which leads to the general conditions of adiabatic evolution: Eqs. (\ref{two_level_cond1}) and (\ref{two_level_cond2}).

\subsection{Non oscillating case}

We assume here that the usual condition (\ref{adia_crit})
 is fulfilled for a 
 strongly non-oscillating real  hamiltonian (i.e. with monotonics 
the $ P_{m k} =  
\langle m^{\mathrm st}  | k\rangle = \varphi_{m k}$ functions) and we give here a clue   that the evolution is indeed adiabatic.

By using a proof by contradiction, we assume that the evolution is not adiabatic. Thus condition (\ref{fourth_th})
is not fulfilled so
 non negligible $\delta'$ off diagonal terms exists to modify substantively the $\delta'$ eigenvalues.
The (Weyl-)Bauer-Fike's theorem (\ref{WBF}) applied to $\delta'$, implies that one of
the off diagonal elements ($ \sim \dot \varphi_{m k}$)  of $\delta'$ should then be bigger than the diagonal ones (the gap $\Delta E_n$).
 But
 Eq. (\ref{adi_cond_1}) indicates that a time $T\sim 1/\| \Omega' \| $ is needed to have an non adiabatic evolution.
 Thus 
 condition (\ref{adia_crit}), which is roughly 
$\|\Omega' \| \ll \Delta E_n $,
 would implies that $\varphi_{m k } \sim \dot \varphi_{m k} T \gg 1 $ which contradicts 
 $\varphi_{m k} = \langle m^{\mathrm st}  | k\rangle  \leq 1$. 
 
 \subsection{(Weyl-)Bauer-Fike's theorem}
 
Let $H'_{\rm d}$ be the diagonal part of $H'=H'_{\rm d}+H'_{\rm non\ diag}$.
 Multiplying 
$H'_{\rm non\ diag} | n' \rangle = (E'_n - H'_{\rm d})    | n' \rangle$ by $(E'_n - H'_{\rm d})^{-1}  $ and taking norm on both sides 
leads to
the (Weyl-Lidskii)-Bauer-Fike's theorem (applied to $H'$): 
\begin{equation}
\min_m |E'_n -  H'_{m m}| \leq \| H'_{\rm non \ diag} \| 
\label{WBF}
\end{equation}

  \subsection{Universal optimal bound}
 
Using the 
following choice
$\theta_m = \int_0^t i \langle m | \dot m \rangle - \int_0^t E_m/\hbar $ of
a geometrical phase (Berry Phase for cyclic evolution) plus a dynamical phase
  simplifies the  $H'$ matrix elements (see
  Eq. \ref{Had_def})).
 Using
 $ - \frac{ d |U_{nn}| }{d t}   \leq  \left|  \frac{ d U_{nn} }{d t} \right|
 $ and the norm\footnoterecall{foot_norm} equality $ \sqrt{1-   |U_{nn}^2| } = \sqrt{ \sum_{m \neq n} | U_{mn}|^2 }  = \| U_{ \bullet  n} \|$, when integrating the
 (Schr\"odinger) equation ($i\hbar \dot U_{nn} = \Omega^\dag U_{ \bullet  n} $)
  leads to the (quantum Zeno's type of)  adiabatic condition:
\begin{equation}
1-|U_{nn} (t) |  \leq 1- \cos (\|\Omega' \|  t/2) \leq \frac{\|\Omega'/2 \|^2}{2} t^2.  
\label{adi_cond_1}
\end{equation}
This optimal bound is reached by the Schwinger  system for $\delta' =0$.

\end{document}